# High temperature behavior of a deformed Fermi gas obeying interpolating statistics

**Abdullah Algin[1] and Mustafa Senay[2]**

[1]*Department of Physics, Eskisehir Osmangazi University, 26480-Eskisehir, Turkey*
[2]*Graduate School of Sciences, Eskisehir Osmangazi University, 26480-Eskisehir, Turkey*

An outstanding idea originally introduced by Greenberg is to investigate whether there is equivalence between intermediate statistics, which may be different from anyonic statistics, and $q$-deformed particle algebra. Also, a model to be studied for addressing such an idea could possibly provide us some new consequences about the interactions of particles as well as their internal structures. Motivated mainly by this idea, in this work, we consider a $q$-deformed Fermi gas model whose statistical properties enable effectively us to study interpolating statistics. Starting with a generalized Fermi–Dirac distribution function, we derive several thermostatistical functions of a gas of these deformed fermions in the thermodynamical limit. We study the high temperature behavior of the system by analyzing the effects of $q$-deformation on the most important thermostatistical characteristics of the system such as the entropy, specific heat, and equation of state. It is shown that such a deformed fermion model in two and three spatial dimensions exhibits the interpolating statistics in a specific interval of the model deformation parameter $0 < q < 1$. In particular, for two and three spatial dimensions, it is found from the behavior of the third virial coefficient of the model that the deformation parameter $q$ interpolates completely between attractive and repulsive systems, including the free boson and fermion cases. From the results obtained in this work, we conclude that such a model could provide much physical insight into some interacting theories of fermions, and could be useful to further study the particle systems with intermediate statistics.

## I. INTRODUCTION

One of the most important theorems in quantum field theory is the spin-statistics theorem, which asserts that particles with spin equal to an even multiple of $\hbar/2$ are bosons and those with spin equal to an odd multiple of $\hbar/2$ are fermions [1-3]. Due to the contrasting nature of symmetry requirements between bosons and fermions, only fermionic particles obey the Pauli exclusion principle. However, studies on non-linear behaviors in complex systems observed in nature indicate sensibly that there could be non-standard type (or types) of quantum statistics outside the standard Bose-Einstein (BE) and Fermi-Dirac (FD) statistics, which therefore could explain the observed non-linearities [4-11]. Some of the earlier examples of non-standard statistics can be mentioned through the works of Gentile [12] and Green [13].

In the last two decades, the two distinct methods in the literature became most popular for studying non-linear behaviors observed in complex systems, which possibly could have non-standard quantum statistical behaviors. Both methods serve as to formulate possible generalizations of statistical mechanics. The first method employs the properties of one- or two-parameter generalized bosonic and fermionic quantum group structures [14-18]. The second method exploits the properties of Tsallis non-extensive statistical mechanics [19,20]. Meanwhile, possible connections between quantum groups and Tsallis non-extensive statistical mechanics have been extensively investigated [21-24].

In studying complex systems, fermionic quantum algebras have found many applications in a wide spectrum of research covering black hole statistics [25], and discussions on some hadronic properties such as the dynamical mass generated for quarks and pure nuclear pairing force version of the Bardeen-Cooper-Schrieffer (BCS) many-body formalism [26,27]. They were also used to understand higher-order effects in the many-body interactions in nuclei [28,29]. By means of quantum algebras, one could also describe interactions between bosons and fermions.

Furthermore, statistical and thermodynamical consequences of studying $q$-deformed physical systems have been intensively investigated in the literature [30-49]. In the framework of $q$-bosons and similar operators, the so-called quons [50], considarable investigations have been carried out for obtaining possible violation of the Pauli exclusion principle [51], and also possible relations to anyonic statistics [52-55]. Also, some recent studies have been used to view $q$-deformations as phenomenological means of introducing an extra parameter, "$q$", to account for non-linearity in the system. Such an approach was considered in [56,57], where some values of $q$ are found to fit the properties of a real (non-ideal) laser and of the gap in the specific heat of a dilute gas of rubidium atoms, respectively. Therefore, such interesting results constitute some motivation to consider quantum algebras for approximating properties of interpolating statistics. Another motivation to consider femionic or bosonic quantum algebras for approaching properties of interpolating statistics comes from the recent developments on some quasiparticle states such as non-Abelian anyons [58], trions [59], and excitons [60].

On the other hand, it was shown in [61-66] that the high and low temperature behaviors of the quantum group covariant bosonic and fermionic oscillator models depend radically on the real deformation parameters. In our recent study [64], we showed that the two-parameter deformed quantum group boson gas exhibits an interesting behavior for high temperatures such that its willingness to show femionlike behavior increases too much for a wide range of the model deformation parameters $q_1$ and $q_2$. Also, the two-parameter deformed quantum group fermion gas model has revealed some notable properties for high temperatures [65]. For instance, such a two-parameter fermion model in two spatial dimensions exhibits a remarkably anyonic type of behavior at some critical values of the model deformation parameters. However, it is impossible to obtain a similar behavior either in the one-parameter deformed $SU_q(2)$-fermion gas model [63] or in the undeformed fermion gas. Moreover, thermodynamical and statistical consequences of the one-parameter deformed bosonic and fermionic oscillator algebras without quantum group symmetries have been recently studied by Lavagno and Narayana Swamy [67-70]. In these works, they have shown that the thermostatistics of one-dimensional $q$-deformed bosonic and fermionic oscillators can be built on the formalism of $q$-calculus. Although, different approximations have been considered to examine the properties of interpolating statistics such as in [71-73], the complete theory of intermediate statistics is currently under active investigation.

Apart from the physical motivations mentioned above, an outstanding idea originally introduced by Greenberg [50,51] is mainly motivated us. His idea was whether there is equivalence between intermediate statistics and $q$-deformed particle algebra with or without quantum group symmetry. Also, a model to be studied for addressing such an idea could possibly provide us some new consequences about the interactions of particles as well as their internal structures. Therefore, in the present work, we will not only pursue this idea, but also we will study on statistical mechanical properties of a fermionic $q$-algebra model, which has some exotic properties such as no exclusion principle. Indeed, it will be shown that the model under consideration enables effectively us to study interpolating statistics, which may indeed be different from that of anyons.

With the above notions in mind, we consider a different generalization of the fermionic system, called Viswanathan-Parthasarathy-Jagannathan-Chaichian (VPJC) oscillators. They have a spectrum given by the *q*-fermionic basic number, and do not satisfy the Pauli exclusion principle when $q \neq 1$. Historically, this model was first introduced by Viswanathan *et al*. [74] and some of its statistical properties were also discussed by Chaichian *et al*. [75]. Therefore, we call them as the VPJC-oscillators model. When compared with the other fermion models in the literature [30-35,63,68-73,76], the VPJC-oscillators model has rather different properties which will be shown below, and it has not fully examined in the past.

Thus, the aim of this paper is twofold. First, we wish to study the high temperature thermostatistical properties of a gas of the VPJC-oscillators. In this sense, we continue to study further the work of [76], and investigate the effects of *q*-deformation on the high temperature behavior of the system. Particular emphasis is given to a discussion on the behaviors of virial coefficients in terms of the deformation parameter *q* in the equation of state for such a deformed fermion system. Second, we want to investigate possible connections between the thermostatistical properties of such a deformed fermion model and the properties of interpolating statistics. Hence, it will be shown below that the model exhibits an interpolation between bosonlike and fermionlike behaviours by means of its third virial coefficient. Therefore, the model is important not only in showing a link between deformed fermions and interpolating statistics, but also in providing a physically interpretable model of *q*-deformed objects.

The paper is organized as follows. In Sec. II, we review the basic algebraic and representative properties of the VPJC-oscillators model. In Sec. III, we investigate the high temperature thermostatistical properties of a gas of such deformed fermion oscillators in the thermodynamical limit. For instance, we derive the distribution function and the equation of state as a virial expansion in the two- and three-dimensional space in order to determine the role of the deformation parameter *q* on the system. In the last section, we discuss possible connections between our results and interpolating statistics, and give our conclusions.

## II. THE VPJC-OSCILLATORS MODEL

In this section, we briefly introduce the basic properties of the VPJC-oscillators model which leads to a suitable framework in order to study the properties of interpolating statistics. The model generated by the VPJC-oscillators $c$ together with their corresponding creation operators $c^*$ is defined by the following relations [74,75]:

$$cc^* + qc^*c = 1 \tag{1}$$
$$[\hat{N}, c] = -c, \quad [\hat{N}, c^*] = c^*,$$

where $\hat{N}$ is the number operator, and *q* is the real positive deformation parameter. Although, this model can be obtained by making a transformation [74,75] from another generalized fermion algebra proposed by Parthasarathy and Viswanathan [34], it notably reveals different quantum algebraic and statistical properties as will be discussed below. Also, when compared with the other *q*-fermion algebras [76], it constitutes less popular model, and it has not fully examined in the literature.

Furthermore, the deformed fermion number operator for the model algebra in Eq. (1) can be obtained as

$$c^*c = [\hat{N}] = \frac{1-(-1)^{\hat{N}}q^{\hat{N}}}{1+q}, \tag{2}$$

whose spectrum is given by

$$[n] = \frac{1-(-1)^n q^n}{1+q}, \tag{3}$$

which is the $q$-fermionic basic number for the VPJC-oscillators model. For $q \neq 1$, one can construct the representations of the operators $c$, $c^*$ in the Fock space spanned on the states $|n\rangle$ of the fermion number operator according to

$$|n\rangle = \frac{(c^*)^n}{\sqrt{[n]!}}|0\rangle, \tag{4}$$

where $[n]! = [n][n-1][n-2]....[1]$. The actions of the operators $c$ and $c^*$ on the Fock states $|n\rangle$ can be obtained as

$$c|n\rangle = \sqrt{[n]}|n-1\rangle, \qquad c^*|n\rangle = \sqrt{[n+1]}|n+1\rangle, \tag{5}$$
$$c^*c|n\rangle = [\hat{N}]|n\rangle = [n]|n\rangle.$$

On the other hand, the Pauli exclusion principle in the VPJC-oscillators model can be recovered only in the limit $q = 1$, since we have

$$\lim_{q \to 1}[1] = 1, \qquad \lim_{q \to 1}[2] = 0, \tag{6}$$

which implies $(c^*)^n = 0$ for $n > 1$ in Eq. (4). Hence, the Fock states for this model reduce to the states $|0\rangle$ and $|1\rangle$. Therefore, we need not to assume the condition $c^2 = 0$ for this model in contrast with the situation introduced in [75]. Also, as pointed out in [74], Eqs. (3) and (4) show that the positive norm condition on the state vectors cannot be satisfied for even values of $n$ if $q > 1$. Therefore, we consider the interval $0 < q < 1$ for the deformation parameter $q$ in this model for the rest of calculations in the present work.

All of the properties mentioned above reveal that the VPJC-oscillators model presents different non-trivial generalized fermions with a spectrum in Eq. (3) without exclusion principle in the interval $0 < q < 1$. It is not only different from the properties of other $q$-fermion models studied by several researchers [30-35,63,68-73,76] but also it differs from the bosonic $q$-oscillator model introduced by Arik and Coon [77]:

$$bb^* - qb^*b = 1, \qquad 0 < q < 1, \tag{7}$$
$$[b,\hat{N}] = b, \qquad [b^*,\hat{N}] = -b^*,$$

whose number operator spectrum was defined by the relation

$$[n] = \frac{1-q^n}{1-q}. \tag{8}$$

Furthermore, following the procedure proposed in [70,78] on another generalized fermion algebra, we now establish the Jackson derivative (JD) [79] appropriate for the model algebra in Eqs. (1)-(3). We may have the holomorphic representation as

$$c \Leftrightarrow D_x, \qquad c^* \Leftrightarrow x. \tag{9}$$

Hence, the model algebra in Eqs. (1)-(3) can be rewritten as

$$D_x x + q x D_x = 1. \tag{10}$$

To derive a solution for this equation, we first observe the following relation

$$x[\hat{N}+1] + q x[\hat{N}] = x, \tag{11}$$

which can be expressed by means of Eq. (5). From Eqs. (10) and (11), and using the property $[\hat{N}]x = x[\hat{N}+1]$, we deduce the following solution:

$$D_x = \frac{1}{x}[\hat{N}] = \frac{1}{x}\left(\frac{1-(-1)^{\hat{N}} q^{\hat{N}}}{1+q}\right). \tag{12}$$

If we use the property $(-q)^{\hat{N}} f(x) = f(-qx)$ [80], this fermionic JD can also be expressed as

$$D_x f(x) = \frac{1}{x}\left(\frac{f(x) - f(-qx)}{1+q}\right), \tag{13}$$

for any function $f(x)$. This fermionic JD does not reduce to the ordinary derivative in the limit $q = 1$. Recently, many of the mathematical properties of this operator were studied by Schork [42]. But, above analysis is different from Schork's study [42], where the author just employed a replacement $q \equiv -\tilde{q}$ with $\tilde{q} > 0$ for the Arik-Coon oscillator algebra in Eqs. (7) and (8). Note that the deformation parameter $q$ for our model algebra in Eqs. (1)-(3) is strictly positive. In particular, if one introduces a function $F(x)$ together with its even part $F_e(x)$ and odd part $F_o(x)$ by

$$F(x) = F_e(x) + F_o(x), \quad F_e(x) = \frac{1}{2}\{F(x) + F(-x)\}, \quad F_o(x) = \frac{1}{2}\{F(x) - F(-x)\}, \tag{14}$$

then one finds that $D_x F_e(x)$ vanishes in the limit $q = 1$, and in the same limit it also reduces to

$$D_x F(x) = \frac{F_o(x)}{x}. \tag{15}$$

A more detailed analysis about the Fock space representations and the fermionic JD on this model was recently carried out in [76]. The fermionic JD in Eq. (13) plays a central role in the framework of mathematical physics such that it is not only needed to define a consistent formulation of the fermionic $q$-calculus, but also it is required to study the thermostatistics of a gas of the VPJC-oscillators.

The fact that when $0 < q < 1$, the VPJC-oscillators model in Eqs. (1)-(3) shows different generalized fermions without exclusion principle, i.e. the Fock states may be occupied by arbitrary number of quanta with $n = 0,1,2,3,....\infty$, could give new interesting results in the framework of statistical mechanics such as unusual realizations of quantum statistics. In the next section, we will particularly focus on the high temperature thermostatistical properties of a gas of the VPJC-oscillators in order to examine the behaviors of intermediate statistics particles.

### III. HIGH TEMPERATURE THERMOSTATISTICS OF THE VPJC-OSCILLATORS

Now we are going to investigate the high temperature (low density) behavior of the VPJC-oscillators gas described by model algebra in Eqs. (1)-(3) in a specific interval $0 < q < 1$. The system containing the VPJC-oscillators constitutes essentially a "free" $q$-deformed fermionic gas system, since the VPJC-oscillators do not interact with each other. The reason behind this consideration is that we do not have both a specific deformed anti-comutation relation between fermionic annihilation (or creation) operators and a quantum group symmetry structure in Eq. (1). In grand canonical ensemble, the model Hamiltonian of such a deformed fermion gas can be expected to have the following form:

$$\hat{H}_F = \sum_i (\varepsilon_i - \mu)\hat{N}_i, \qquad (16)$$

where $\varepsilon_i$ is the kinetic energy of a particle in the $i$ state, and $\mu$ is the chemical potential. Similar Hamiltonians were also considered by several other authors [30-41,67-73,75,76]. We should concisely point out that the form in Eq. (16) shows a deformed Hamiltonian, which depends implicitly on the deformation parameter $q$. Since we have the number operator defined in Eq. (2). The mean value of the $q$-deformed occupation number $n_i$ is defined by [37]

$$[n_i] = \frac{1}{Z}Tr(e^{-\beta\hat{H}_F}[\hat{N}_i]) \equiv \frac{1}{Z}Tr(e^{-\beta\hat{H}_F}c_i^*c_i), \qquad (17)$$

where $\beta = 1/kT$, $k$ is the Boltzmann constant, $T$ is the temperature of the system, and $Z = Tr(e^{-\beta\hat{H}_F})$ is the partition function. After applying the cyclic property of the trace [30], and using the Fock space properties of the VPJC-oscillators algebra in Eqs. (1) and (5), we obtain

$$\frac{[n_i]}{[n_i+1]} = e^{-\beta(\varepsilon_i-\mu)}. \qquad (18)$$

Also, from Eqs. (3) and (6), we have

$$\lim_{q \to 1} [n] = n, \quad \lim_{q \to 1} [n+1] = 1 - n, \quad n = 0, 1, \tag{19}$$

where we have dropped the subscript *i* for the sake of simplicity. Therefore, the expression in Eq. (18) reduces to the usual Fermi-Dirac distribution in the limit $q = 1$. From the definition of the *q*-fermion basic number $[n]$ in Eq. (3) and using Eq. (18), we derive

$$n = n(\eta, q) = \frac{1}{|\ln q|} \left| \ln \left( \frac{|e^\eta - 1|}{e^\eta + q} \right) \right|, \tag{20}$$

where $\eta = \beta(\varepsilon - \mu)$, and the deformation parameter *q* has values in the interval $0 < q < 1$. This equation provides the *q*-fermion distribution of the VPJC-oscillators model, which may also be expressed as statistical distribution function for a gas of the particles obeying interpolating statistics. Moreover, it satisfies the positivity condition to be the correct fermion distribution function. Also, one should consider the relations in Eqs. (18) and (19) in order to find the usual Fermi-Dirac distribution as $n(\eta) = 1/(e^\eta + 1)$, which has a similar modified form for the case $q = 1$ for finite temperatures. It takes the standard step-functional form in the limit $T = 0$ for any values of *q*. Hence, we conclude that the *q*-deformation of fermions is just a finite temperature effect in the present VPJC-oscillators model. It is discontinuous at $\varepsilon = \mu$ and also, the peak of this function occurs at $\eta = 0$ for any values of *q*. In Fig. 1, the *q*-deformed statistical distribution function $n(\eta, q)$ of the VPJC-oscillators is shown for finite temperatures as a function of $\eta = \beta(\varepsilon - \mu)$ for values of the deformation parameter *q* smaller than 1. Consequently, the behavior of the statistical distribution for interpolating statistics particles exhibited by the present fermion model is rather different from both the usual fermion distribution and the other *q*-fermion models considered in the literature [76].

Using the relations in Eqs. (13) and (20), we deduce the logarithm of the fermionic grand partition function as

$$\ln Z_F = \frac{(1+q)}{|\ln q|} \sum_i \left| \ln \left| (1 - z e^{-\beta \varepsilon_i}) \right| \right|, \tag{21}$$

where $0 < q < 1$, and the fugacity is assumed to have the standard form $z = e^{\beta \mu}$. However, the standard thermodynamic relations in the usual form are ruled out. For instance, the total number particles in the VPJC-oscillators gas model cannot be obtained by using a standard thermodynamical expression such as

$$N \neq z \left( \frac{\partial}{\partial z} \right) \ln Z_F. \tag{22}$$

Here, an important point is to use the fermionic JD in Eq. (13) instead of the usual thermodynamics derivative with respect to *z* as follows:

$$\frac{\partial}{\partial z} \to D_z^{(q)}, \tag{23}$$

where the fermionic JD $D_z^{(q)}$ has the same form as in Eq. (13) with the variable $z$. Therefore, the total number of particles in the VPJC-oscillators gas can be derived from the relation

$$N = zD_z^{(q)} \ln Z_F \equiv \sum_i n_i \,, \tag{24}$$

where $n_i$ is expressed by Eq. (20). In order to obtain the high temperature thermostatistical characteristics of the system, we can replace the sums over states by integrals for a large volume and a large number of particles [81-84]. Accordingly, the equation of state $(PV/kT) = \ln Z_F$ can be written as

$$\frac{P}{kT} = \frac{(1+q)}{|\ln q|} \frac{4\pi}{h^3} \int_0^\infty dp\, p^2 \left| \ln \left| (1 - ze^{-\beta p^2/2m}) \right| \right|, \tag{25}$$

where $0 < q < 1$. Similary, the particle density for the VPJC-oscillators is

$$\frac{1}{\upsilon} = \frac{N}{V} = \frac{1}{|\ln q|} \frac{4\pi}{h^3} \int_0^\infty p^2 dp \left| \ln \left( \frac{\left|(1 - ze^{-\beta p^2/2m})\right|}{1 + qze^{-\beta p^2/2m}} \right) \right|. \tag{26}$$

Using the fermionic JD in Eq. (13) and after straightforward manipulation, Eqs. (25) and (26) can be rewritten as

$$\frac{P}{kT} = \frac{1}{\lambda^3} f_{5/2}(z,q) \,, \tag{27}$$

$$\frac{1}{\upsilon} = \frac{1}{\lambda^3} f_{3/2}(z,q) \,, \tag{28}$$

where $\lambda = \sqrt{2\pi\hbar^2/mkT}$ is the thermal wavelength, and the one-parameter generalized Fermi-Dirac function $f_n(z,q)$ is defined as

$$f_n(z,q) = \frac{1}{\Gamma(n)} \int_0^\infty x^{n-1} dx \frac{1}{|\ln q|} \left| \ln \left( \frac{\left|(1-ze^{-x})\right|}{1+qze^{-x}} \right) \right|$$

$$= \frac{1}{|\ln q|} \left( \sum_{l=1}^\infty (-1)^{l-1} \frac{(zq)^l}{l^{n+1}} - \sum_{l=1}^\infty \frac{z^l}{l^{n+1}} \right), \tag{29}$$

where $x = \beta\varepsilon$ and $\varepsilon = p^2/2m$. This one-parameter generalized function reduces to the standard Fermi-Dirac functions $f_n(z)$ in the limit $q=1$, when we consider the discussions made after Eq. (20). Moreover, these functions are different from the functions $h(n,z,q)$ of [68-70]. In order to compare the behaviors of the $q$-deformed functions $f_n(z,q)$ and the standard functions $f_n(z)$, in Figs. 2 and 3, the plots of these functions are shown as a function of $z$ for several values of the deformation parameter $q$ for the case $q<1$, respectively. When we compare with the $q = 1$ case in Fig. 3, the values of the $q$-deformed Fermi-Dirac functions $f_{3/2}(z,q)$ and $f_{5/2}(z,q)$ for the case $q<1$ increase with the value of the deformation parameter $q$ as shown in Fig. 2. Also, the values of these deformed functions for $q<1$ are larger than the standard $f_n(z)$ functions at the same fugacity. According to Figs. 2 and 3, the values of the function $f_{3/2}(z,q)$ are larger than those of the function $f_{5/2}(z,q)$ for the case $q<1$ in contrast to the behaviors of the standard functions $f_{3/2}(z)$ and $f_{5/2}(z)$.

The method proposed in [68] can be applied to find the internal energy of the VPJC-oscillators gas in the present model. In this calculation, we consider the prescription for the fermionic JD in Eq. (13) and the ordinary chain rule as follows:

$$U = \left(-\frac{\partial \ln Z_F}{\partial \beta}\right) = -\frac{(1+q)}{|\ln q|} \sum_i \frac{\partial y_i}{\partial \beta} D_{y_i}^{(q)} \left\| \ln |(1-zy_i)| \right\| , \qquad (30)$$

where $y_i = \exp(-\beta \varepsilon_i)$. This equation leads to

$$U = \sum_i \varepsilon_i n_i , \qquad (31)$$

where $n_i$ is expressed by Eq. (20). We can also obtain the internal energy as

$$\frac{U}{V} = \frac{3}{2} \frac{kT}{\lambda^3} f_{5/2}(z,q) . \qquad (32)$$

From Eqs. (13), (21), (28), (32), the entropy for the VPJC-oscillators gas can be found as

$$\frac{S^q}{Nk} = \frac{5}{2} \frac{f_{5/2}(z,q)}{f_{3/2}(z,q)} - \ln z . \qquad (33)$$

For comparison, in Figs. 4 and 5, we present the plots of the $q$-deformed entropy function $S^q/Nk$ and the entropy function $S/Nk$ of a free fermion gas as a function of $z$ for values of the deformation parameter $q$ for the cases $q<1$ and $q=1$, respectively. For $q<1$, the entropy values of the VPJC-oscillator gas decrease with the values of the deformation parameter $q$. Also, the entropy values of the VPJC-oscillators gas for the interval $0<q<1$ are lower than the results of a free fermion gas at the same fugacity as shown in Figs. 4 and 5.

With the above results in mind, the specific heat of the VPJC-oscillators gas can be obtained from the thermodynamic definition $C_V = (\partial U/\partial T)_{V,N}$. For high temperatures and making use of the fermionic JD in Eq. (13), the specific heat of our model is

$$\frac{C_V^q \lambda^3}{kV} = \frac{15}{4} z D_z f_{7/2}(z,q) - \frac{9}{4} z \frac{(D_z f_{5/2}(z,q))^2}{D_z f_{3/2}(z,q)} , \qquad (34)$$

which has rather different form than that of a free fermion gas [82].

On the other hand, for high temperatures, i.e. the limit $\lambda^3/\upsilon \ll 1$, using Eqs. (28) and (29), we find the fugacity $z$ in terms of the deformation parameter $q$ up to the fourth order in $(\lambda^3/\upsilon)$ as

$$z = \frac{|\ln q|}{(q-1)}\left(\frac{\lambda^3}{\upsilon}\right) + \frac{1}{2^{5/2}}\frac{|\ln q|^2}{(q-1)^3}(q^2+1)\left(\frac{\lambda^3}{\upsilon}\right)^2$$
$$+ \left[\frac{1}{2^4}\frac{|\ln q|^3}{(q-1)^5}(q^2+1)^2 - \frac{1}{3^{5/2}}\frac{|\ln q|^3}{(q-1)^4}(q^3-1)\right]\left(\frac{\lambda^3}{\upsilon}\right)^3 \quad (35)$$
$$+ \left[\frac{5}{2^{15/2}}\frac{|\ln q|^4}{(q-1)^7}(q^2+1)^3 - \frac{5}{6^{5/2}}\frac{|\ln q|^4}{(q-1)^6}(q^2+1)(q^3-1) + \frac{1}{2^5}\frac{|\ln q|^4}{(q-1)^5}(q^4+1)\right]\left(\frac{\lambda^3}{\upsilon}\right)^4,$$

where $0 < q < 1$. The equation of state for the VPJC-oscillators system can also be derived from Eqs. (27) and (28) as a virial expansion in the three-dimensional space:

$$\frac{PV}{NkT} = \sum_{l=1}^{\infty} a_l(q) \left(\frac{\lambda^3 N}{V}\right)^{l-1}, \quad (36)$$

where the first few virial coefficients $a_l(q)$ are given by

$$a_1(q) = 1, \quad (37)$$

$$a_2(q) = \frac{1}{2^{7/2}}\frac{|\ln q|}{(q-1)^2}(q^2+1), \quad (38)$$

$$a_3(q) = \frac{1}{2^5}\frac{|\ln q|^2}{(q-1)^4}(q^2+1)^2 - \frac{2}{3^{7/2}}\frac{|\ln q|^2}{(q-1)^3}(q^3-1), \quad (39)$$

$$a_4(q) = \frac{5}{2^{17/2}}\frac{|\ln q|^3}{(q-1)^6}(q^2+1)^3 - \frac{3}{6^{5/2}}\frac{|\ln q|^3}{(q-1)^5}(q^3-1)(q^2+1) + \frac{3}{2^7}\frac{|\ln q|^3}{(q-1)^4}(q^4+1), \quad (40)$$

where $0 < q < 1$. We should note that the first coefficient is exact, since it does not contain any term with deformation parameter $q$. The other virial coefficients contain the higher order terms depending on the several powers of the deformation parameter $q$, which may be considered as the deviations from the corresponding standard values of these coefficients for a free fermion gas system [82]. Since we are dealing with the high temperature limit, we are particularly focusing our attention to the virial coefficients $a_2(q)$, $a_3(q)$, $a_4(q)$ and omitting the other terms. Obviously, the signs of these virial coefficients depend on the values of the deformation parameter $q$. Therefore, this parameter is responsible for the behavior of the present fermion gas model. Figure 6 shows a graph of these virial coefficients as a function of the deformation parameter $q$ for the interval $0 < q < 1$.

On the other hand, it is interesting to investigate whether a similar behavior could be found for the two-dimensional system by performing the same calculations. If one follows the same procedure as above, then the equation of state can be derived as

$$\frac{PA}{NkT} = \sum_{l=1}^{\infty} \tilde{a}_l(q) \left( \frac{\lambda^2 N}{A} \right)^{l-1}, \tag{41}$$

where $A$ is the surface confining the fermionic system, and the first few virial coefficients $\tilde{a}_l(q)$ in the two-dimensional space are as follows:

$$\tilde{a}_1(q) = 1, \tag{42}$$

$$\tilde{a}_2(q) = \frac{1}{2^3} \frac{|\ln q|}{(q-1)^2} (q^2 + 1), \tag{43}$$

$$\tilde{a}_3(q) = \frac{1}{2^4} \frac{|\ln q|^2}{(q-1)^4} (q^2 + 1)^2 - \frac{2}{3^3} \frac{|\ln q|^2}{(q-1)^3} (q^3 - 1), \tag{44}$$

$$\tilde{a}_4(q) = \frac{5}{2^7} \frac{|\ln q|^3}{(q-1)^6} (q^2 + 1)^3 - \frac{1}{2^2 3} \frac{|\ln q|^3}{(q-1)^5} (q^3 - 1)(q^2 + 1) + \frac{3}{2^6} \frac{|\ln q|^3}{(q-1)^4} (q^4 + 1), \tag{45}$$

where $0 < q < 1$, and the first coefficient is again exact since it does not contain any term with the deformation parameter $q$. The other virial coefficients contain the higher order terms depending on the several powers of the deformation parameter $q$, which may also be considered as the deviations from the corresponding standard values of these coefficients for a free fermion gas system in two dimensions [7]. As in the above three-dimensional case, we focus on the behaviors of the virial coefficients in Eqs. (43)-(45) in order to find the effects of $q$-deformation for high temperatures, and we omit the other virial coefficients in Eq. (41). In this way, we wish to find out how the $q$-deformation effects the high temperature quantum statistical behavior of the present fermion model. Figure 7 shows a graph of these three virial coefficients as a function of the deformation parameter $q$ for the interval $0 < q < 1$. Obviously, the signs of the virial coefficients in Eqs. (43)-(45) depend on the values of the deformation parameter $q$.

We wish to close this section by particularly discussing the third virial coefficients $a_3(q)$ and $\tilde{a}_3(q)$ for both the three- and two-dimensional systems in Eqs. (39) and (44). It is seen from Figs. 6 and 7, the values of the coefficients are always positive for the interval $0 < q < 1$ except that the virial coefficients $a_3(q)$ and $\tilde{a}_3(q)$. The signs of the third virial coefficients in Eqs. (39) and (44) change depending on the values of the deformation parameter $q$. For instance, in the three-dimensional space, the third virial coefficient $a_3(q)$ has just negative values for high temperatures for the case $q \leq 0.2019$ as shown in Fig. 6. On the other hand, in the two-dimensional case, the third virial coefficient $\tilde{a}_3(q)$ has just negative values for the case $q \leq 0.1270$ as shown in Fig. 7. Therefore, these third virial coefficients are responsible for the behavior of the present fermion model. The free fermion gas results $a_3(q) = 0.0033$ [83] and $\tilde{a}_3(q) = 0.0278$ [7] are reached at $q = 0.2091$ and $q = 0.1632$, respectively. However, the free boson gas results $a_3(q) = -0.0033$ [83] and $\tilde{a}_3(q) = -0.0278$ [7] are reached at $q = 0.1947$ and $q = 0.0948$, respectively. Indeed, this remarkable point is essential difference between the present femionic gas model and the earlier one- and two-parameter deformed fermionic gas models [61-66,76]. Although, it seems to be of minor importance in the high temperature limit, when compared to the other virial coefficients in Eqs. (38), (40), and (43), (45); according to our point of view it brings about a separation between attractive ($q \leq 0.2019$ or $q \leq 0.1270$) and repulsive ($q > 0.2019$ or $q > 0.1270$) behaviors of the system in two and three spatial dimensions.

Therefore, as far as we concern the third virial coefficients $a_3(q)$ and $\tilde{a}_3(q)$ in the high temperature limit, the original fermionic character of the VPJC-oscillators gas model would change to a bosonlike behavior for those values of the deformation parameter $q$ in the cases $q \leq 0.2019$ and $q \leq 0.1270$, respectively. The VPJC-oscillators gas model exhibits bosonlike behaviors in that critical values of the deformation parameter $q$. Such type of behavior can typically be seen in the case of anyonic gas systems. Since, the VPJC-oscillators model interpolates between bosonlike and fermionic behaviors up to the third virial coefficients in two dimensions, we can find a relation between the deformation parameter $q$ of our model and the statistics determining parameter $\alpha$ for an anyon gas [85] in two spatial dimensions. Accordingly, we obtain the following relation:

$$\alpha = \alpha(q) = \left[\frac{1}{2} + \frac{1}{4\sqrt{3}} \frac{|\ln q|}{(q-1)^2} \sqrt{27(q^2+1)^2 - 32(q^3-1)(q-1)}\right]^{1/2}, \quad (46)$$

where $q < 1$.

By considering the above results, we should emphasize that the results in Eqs. (20)-(46) are not only different from the results for the one- and two-parameter fermion models studied in [61-66,76], but also they could serve to present willingness of bosonlike behavior of the VPJC-oscillators model through some critical values of the deformation parameter $q$. Therefore, the deformation parameter $q$ serves as interpolating object between repulsive and attractive behaviors of the system as are shown in Figs. 6 and 7. We conclude that all above considerations give main reasons for considering the VPJC-oscillators as interpolating statistics objects.

However, it must be pointed out for all equations above that the free fermion gas results can only be recovered upon recognizing the discussions after Eq. (20) and applying the limit $q = 1$. In the next section, the other effects of the deformation parameter $q$ on the high temperature thermostatistical behavior of the VPJC-oscillators gas will be discussed.

## IV. DISCUSSION AND CONCLUSION

In this paper, we studied the high temperature behavior of a deformed fermion gas model. Starting with a $q$-deformed Fermi-Dirac distribution function and with the use of the fermionic JD, we calculated various thermostatistical functions of the VPJC-oscillators gas model, and consequently the equation of state is obtained as a virial expansion in the two- and three-dimensional space. In this context, the first four virial coefficients are examined for the interval $0 < q < 1$. We found particularly that the signs of the third virial coefficients $a_3(q)$ and $\tilde{a}_3(q)$ depend on the deformation parameter $q$ in both two and three spatial dimensions. According to Figs. 6 and 7, the deformation parameter $q$ interpolates between bosonlike and fermionic behaviors. Also, in both two and three dimensions, we may remark that the VPJC-oscillators gas model exhibits an interpolation between attractive and repulsive systems for some critical values of the deformation parameter $q$ as indicated above. Therefore, the present fermion model presents a different system containing interpolating statistics particles such that the deformation parameter $q$ may also be interpreted as an interpolating object between fermionic and bosonic characters of the system. When we compared with the high temperature thermostatistical properties of both other $q$-fermion gases [76] and the quantum group symmetric fermion gases [61-66], different properties of the VPJC-oscillators given in Eqs. (20)-(46) enable us to call the model as obeying the interpolating statistics, which was not shown before by other $q$-fermion models in this field of research. As far as we know from the literature, this is the first attempt to introduce the interpolating statistics from a deformed fermion gas model having the properties in Eqs. (1)-(6), (9)-(15). This model in two and three spatial dimensions has a crucial behavior through the deformation parameter $q$. For instance, as far as we concern on the third virial coefficient $\tilde{a}_3(q)$ in two dimensions, it exhibits an interpolation between attractive ($q \leq 0.1270$) and repulsive ($q > 0.1270$) systems including the free boson and free fermion cases as shown in Fig. 7. Hence, we have shown that this simple VPJC-fermionic system describes such kinds of different systems spanned from bosonlike to fermionic regions. Such a result indicates a similar physical behavior as seen in anyonic systems [52-55], whereas apart from the recent study of Lavagno and Narayana Swamy [86], it was not fully examined yet such a behavior in the case of other deformed fermion gases in two spatial dimensions. It should be noted that the $q$-fermion algebra studied in [86] is different from the present VPJC-oscillators model. As a result, this study also shows us that a deformed fermion model can lead to interpolating statistics behavior without the use of any quantum group covariance. In this sense, we may remark that the properties of interpolating statistics could really different from the properties of anyonic systems with fractional statistics. However, our results are in contrast to the consequences of the work of Narayana Swamy [70], where he exploited a different definition of the one-dimensional $q$-fermionic algebra.

On the other hand, the other effects of $q$-deformation on the high temperature thermostatistical properties of the system can be summarized as follows: (i) According to Fig. 1, the values of the $q$-deformed Fermi-Dirac distribution function $n(\eta,q)$ of the VPJC-oscillators model for the case $q<1$ increase when the deformation parameter $q$ is increased at the same $\eta$ values. Also, the $q$-deformed Fermi-Dirac distribution function in Eq. (20) takes the standard step-functional form in the limit $T = 0$. However, the VPJC-oscillators do not satisfy the Pauli exclusion principle for the interval $0<q<1$. In this sense, we can say for instance that it is possible to occupy more than two $q$-fermions in a given quantum state. Such a consideration implies that the present fermion oscillators behave like bosons for those values of $q$, and hence they could lead to supersymmetrization in a way that we can collect them together in the same Fock states of SUSY generators. (ii) According to Fig. 2, the values of the deformed functions $f_n(z,q)$ for $q<1$ increase when the deformation parameter $q$ is increased. However, such a result is in contrast to the behavior of corresponding deformed $h(n,z,q)$ functions of another $q$-fermion model studied in [70]. (iii) According to Figs. 4 and 5, the entropy of the VPJC-oscillators gas decreases for $q<1$ such that its minimum value occurring at $q = 0.99$ is lower than the values of corresponding entropy functions of both a free fermion gas and another $q$-fermion model [70] at the same fugacity. (iv) We should emphasize that the values of all deformed thermostatistical functions have more sensitive variations to those $q$ values, which are in a specific interval $0.8 \leq q <1$. For this range, except that the entropy of the present model, all deformed functions considered increase radically.

Furthermore, from Eqs. (38)-(40) and (43)-(45), we may alternatively interpret the effect of $q$-deformation on the thermostatistics of the model for both two- and three-dimensional space for high temperatures as follows. In particular, when we focus on the behavior of the third virial coefficients shown in Figs. 6 and 7, they change their signs according to some specific values of the deformation parameter $q$, which may be regarded as a control parameter to interpolate between fermionic and bosonlike systems. It seems that the interactions among deformed fermions are such that they lead to a behavior similar to bosonic character for the interpolating statistics particles of the present model. Therefore, both the higher order contributions in Eqs. (39) and (44) and the nature of the signs of these coefficients deserve an interpretation, so that the representation used here brings about an interacting system of fermionlike particles obeying interpolating statistics. Hence, all above considerations indicate us the presence of such an interaction in the system. A parallel discussion was made by the works of Scarfone and Narayana Swamy [87,88] with the use of a $q$-bosons system. They investigated the possibility of finding whether $q$-deformation can be originated from the interactions among bosons through a form of the equation of state in the $q$-bosons system for three dimensions [87,88].

As a consequence, our results in this study can be hopefully applied to approximate nonlinear behaviors of other realistic systems such as cluster expansions for either real gases or interacting fluids and understanding the higher order interaction terms in many-body quantum systems.

The low temperature behavior of the present one-parameter fermion model and other aspects of interpolating statistics are some open problems, parallel to this study to be pursued in the near future.

**ACKNOWLEDGMENTS**

We would like to thank the referees for their useful comments.

[1] W. Pauli, Phys. Rev. **58**, 716 (1940).
[2] S. Weinberg, Phys. Rev. **133**, B1318 (1964).
[3] R. Shankar, *Principles of Quantum Mechanics* (Kluwer Academic, New York, 1994).
[4] J. M. Leinaas, and J. Myrheim, Nuovo Cimento B **37**, 1 (1977).
[5] F. D. M. Haldane, Phys. Rev. Lett. **67**, 937 (1991).
[6] S. Forte, Rev. Mod. Phys. **64**, 193 (1992).
[7] A. Khare, *Fractional Statistics and Quantum Theory* (World Scientific, Singapore, 2005).
[8] R. M. May, Phys. Rev. **135**, A1515 (1964).
[9] W. A. Perkins, Int. J. Theor. Phys. **41**, 823 (2002).
[10] G. A. Goldin and D. H. Sharp, Phys. Rev. Lett. **76**, 1183 (1996).
[11] S. Chaturvedi and V. Srinivasan, Physica A **246**, 576 (1997).
[12] G. Gentile, Nuovo Cimento **17**, 493 (1940).
[13] H. S. Green, Phys. Rev. **90**, 270 (1953).
[14] V. G. Drinfeld, *Proceedings of the International Congress of Mathematics*, *Berkeley* (American Mathematical Society, Providence, RI, (1987), Vol.1, p. 798.
[15] M. Jimbo, Lett. Math . Phys. **11**, 247 (1986).
[16] L. D. Faddeev, N. Y. Reshetikhin, and L. A. Takhtajan, Algebr. Anal. **1**, 129 (1988).
[17] L. C. Biedenharn, J. Phys. A: Math. Gen. **22**, L873 (1989).
[18] A. J. Macfarlane, J. Phys. A: Math. Gen. **22**, 4581 (1989).
[19] C. Tsallis, J. Stat. Phys. **52**, 479 (1988).
[20] E. M. F. Curado and C. Tsallis, J. Phys. A: Math. Gen. **24**, L69 (1991).
[21] C. Tsallis, Phys. Lett. A **195**, 329 (1994).
[22] S. Abe, Phys. Lett. A **224**, 326, (1997).
[23] R. S. Johal, Phys. Rev. E **58**, 4147 (1998).
[24] S. F. Ozeren, U. Tirnakli, F. Buyukkilic, and D. Demirhan, Eur. Phys. J. B **2**, 101 (1998).
[25] A. Strominger, Phys. Rev. Lett. **71**, 3397 (1993).
[26] L. Tripodi, C. L. Lima, Phys. Lett. B **412**, 7 (1997).
[27] V. S. Timoteo, C. L. Lima, Phys. Lett. B **448**, 1 (1999).

[28] K. D. Sviratcheva, C. Bahri, A. I. Georgieva, J. P. Draayer, Phys. Rev. Lett. **93**, 152501 (2004).
[29] A. Ballesteros, O. Civitarese, F. J. Herranz, M. Reboiro, Phys. Rev. C **66**, 064317 (2002).
[30] C. R. Lee and J. P. Yu, Phys. Lett. A **150**, 63 (1990); C. R. Lee and J. P. Yu, Phys. Lett. A **164**, 164 (1992).
[31] M. L. Ge, G. Su, J. Phys. A; Math. Gen. **24**, L721 (1991).
[32] M. Chaichian, D. Ellinas, P. Kulish, Phys, Rev, Lett. **65**, 980 (1990).
[33] Y. J. Ng, J. Phys. A; Math. Gen. **23**, 1023 (1990).
[34] R. Parthasarathy, K. S. Viswanathan, J. Phys. A: Math. Gen. **24**, 613 (1991).
[35] H. S. Song, S. X. Ding, and I. An, J. Phys. A: Math. Gen. **26**, 5197 (1993).
[36] G. Kaniadakis, A. Lavagno, and P. Quarati, Phys. Lett. A **227**, 227 (1997).
[37] J. A. Tuszynski, J. L. Rubin, J. Meyer, and M. Kibler, Phys. Lett. A **175**, 173 (1993).
[38] T. Altherr, T. Grandou, Nucl. Phys. B **402**, 195 (1993).
[39] M. R. Monteiro, L. M. C. S. Rodrigues, and S. Wulck, Phys. Rev. Lett. **76**, 1098 (1996).
[40] S. Cai, G. Su, and J. Chen, J. Phys. A: Math. Theor. **40**, 11245 (2007).
[41] S. Cai, G. Su, and J. Chen, Int. J. Mod. Phys. B **24**, 3323 (2010).
[42] M. Schork, Russ. J. Math. Phys. **12**, 394 (2005).
[43] Q. Zeng, Z. Cheng, and J. Yuan, J. Stat. Mech.: Theor. Exp. P11035 (2010).
[44] A. M. Gavrilik, I. I. Kachurik, and Y. U. Mishchenko, J. Phys. A: Math. Theor. **44**, 475303 (2011).
[45] A. M. Gavrilik and A. P. Rebesh, Mod. Phys. Lett. B **26**, 1150030 (2012).
[46] A. P. Polychronakos, Phys. Lett. B **365**, 202 (1996).
[47] B. Mirza and H. Mohammadzadeh, J. Phys. A: Math. Theor. **44**, 475003 (2011).
[48] D. Bonatsos and C. Daskoloyannis, Prog. Part. Nucl. Phys. **43**, 537 (1999).
[49] S. L. Dalton, A. Inomata, Phys. Lett. A **199**, 315 (1995).
[50] O. W. Greenberg, Phys. Rev. Lett. **64**, 705 (1990).
[51] O. W. Greenberg, Phys. Rev. D **43**, 4111 (1991).
[52] F. Wilczek, Phys. Rev. Lett. **48**, 1144 (1982); F. Wilczek, Phys. Rev. Lett. **49**, 957 (1982).
[53] D. Arovas, R. Schrieffer, F. Wilczek, and A. Zee, Nucl. Pyhs. B **251**, 117 (1985).
[54] D. Fivel, Phys. Rev. Lett. **65**, 3361 (1990).
[55] M. Frau, A. Lerda, and S. Sciuto, e-print arxiv: hep-th/9407161v1.
[56] J. Katriel and A. I. Solomon, Phys. Rev. A **49**, 5149 (1994).
[57] A. Algin and E. Arslan, J. Phys. A: Math. Theor. **41**, 365006 (2008).
[58] R. L. Willett, L. N. Pfeiffer, and K. W. West, Phys. Rev. B **82**, 205301 (2010).
[59] R. Matsunaga, K. Matsuda, and Y. Kanemitsu, Phys. Rev. Lett. **106**, 037404 (2011).
[60] M. B. Harouni, R. Roknizadeh, and M.H. Naderi, J. Phys. B: At. Mol. Opt. Phys. **42**, 095501 (2009).
[61] M. R. Ubriaco, Phys. Lett. A **219**, 205 (1996).
[62] M. R. Ubriaco, Mod. Phys. Lett A **11**, 2325 (1996).
[63] M. R. Ubriaco, Phys. Rev. E **55**, 291 (1997).
[64] A. Algin, Phys. Lett. A **292**, 251 (2002).
[65] A. Algin, M. Arik, and A. S. Arikan, Phys. Rev. E **65**, 026140 (2002).
[66] M. Arik and J. Kornfilt, Phys. Lett. A **300**, 392 (2002).
[67] A. Lavagno and P. Narayana Swamy, Phys. Rev E **61**, 1218 (2000).
[68] A. Lavagno and P. Narayana Swamy, Phys. Rev E **65**, 036101 (2002).
[69] A. Lavagno and P. Narayana Swamy, Found. Phys. **40**, 814 (2010).
[70] P. Narayana Swamy, Int. J. Mod. Phys. B **20**, 2537 (2006).
[71] R. Acharya and P. Narayana Swamy, J. Phys. A: Math. Gen. **37**, 2527 (2004); R. Acharya and P. Narayana Swamy, J. Phys. A: Math. Gen. **37**, 6605 (2004).

[72] P. Narayana Swamy, Int. J. Mod. Phys. B **20**, 697 (2006); e-print arxiv: quant-ph/0412189vl.
[73] A. Lavagno and P. Narayana Swamy, Physica A **389**, 993 (2010).
[74] K. S. Viswanathan, R. Parthasarathy, R. Jagannathan, J. Phys. A: Math. Gen. **25**, L335 (1992).
[75] M. Chaichian, R. G. Felipe, C. Montonen, J. Phys. A: Math. Gen. **26**, 4017 (1993).
[76] A. Algin, Int. J. Theor. Phys. **50**, 1554 (2011).
[77] M. Arik and D. D Coon, J. Math. Phys. **17**, 524 (1976).
[78] P. Narayana Swamy, Eur. Phys. J. B **50**, 291 (2006).
[79] F. Jackson, Messenger Math. **38**, 57 (1909).
[80] P. Narayana Swamy, Physica A **328**, 145 (2003).
[81] K. Huang, *Statistical Mechanics* (Wiley, New York, 1987).
[82] W. Greiner, L. Neise, H. Stöcker, *Thermodynamics and Statistical Mechanics* (Springer-Verlag, New York, 1994).
[83] R. K. Pathria, *Statistical Mechanics* (Butterworth-Heinemann, New York, 1996).
[84] L. E. Reichl, *A Modern Course in Statistical Physics* (Wiley, New York, 1998).
[85] R. Acharya and P. Narayana Swamy, J. Phys. A: Math. Gen. **27**, 7247 (1994).
[86] A. Lavagno and P. Narayana Swamy, Int. J. Mod. Phys. B **23**, 235 (2009).
[87] A. M. Scarfone and P. Narayana Swamy, J. Phys. A: Math. Theor. **41**, 275211 (2008).
[88] A. M. Scarfone and P. Narayana Swamy, J. Stat. Mech.: Theor. Exp. P02055 (2009).

## LIST OF THE FIGURE CAPTIONS

**FIG. 1.** The $q$-deformed Fermi-Dirac distribution $n(\eta, q)$ as a function of $\eta = \beta(\varepsilon - \mu)$ for values of the deformation parameter $q$ smaller than 1 for finite temperatures.

**FIG. 2.** (Color online) The $q$-deformed Fermi-Dirac function $f_n(z, q)$ (red dotted line, $n = 5/2$; black dashed line, $n = 3/2$) as a function of $z$ for the case $q < 1$.

**FIG. 3.** The standard Fermi-Dirac function $f_n(z)$ (solid line, $n = 5/2$; dashed line, $n = 3/2$) as a function of $z$.

**FIG. 4.** The $q$-deformed entropy function $S^q/Nk$ as a function of $z$ for the case $q < 1$.

**FIG. 5.** The entropy function $S/Nk$ for a free fermion gas as a function of $z$.

**FIG. 6.** (Color online) The virial coefficients $a_l(q)$ (blue dotted line, $n = 4$; black solid line, $n = 3$; and red dashed line, $n = 2$) as a function of the deformation parameter $q$ for a three-dimensional system. The line at $q = 0.2019$ separates the region between $a_3(q) < 0$ and $a_3(q) > 0$, which corresponds to bosonlike and fermionic behavior, respectively.

**FIG. 7.** (Color online) The virial coefficients $\tilde{a}_l(q)$ (blue dotted line, $n = 4$; black solid line, $n = 3$; and red dashed line, $n = 2$) as a function of the deformation parameter $q$ for a two-dimensional system. The line at $q = 0.1270$ divides the region between $\tilde{a}_3(q) < 0$ and $\tilde{a}_3(q) > 0$, which corresponds to bosonlike and fermionic behavior, respectively.

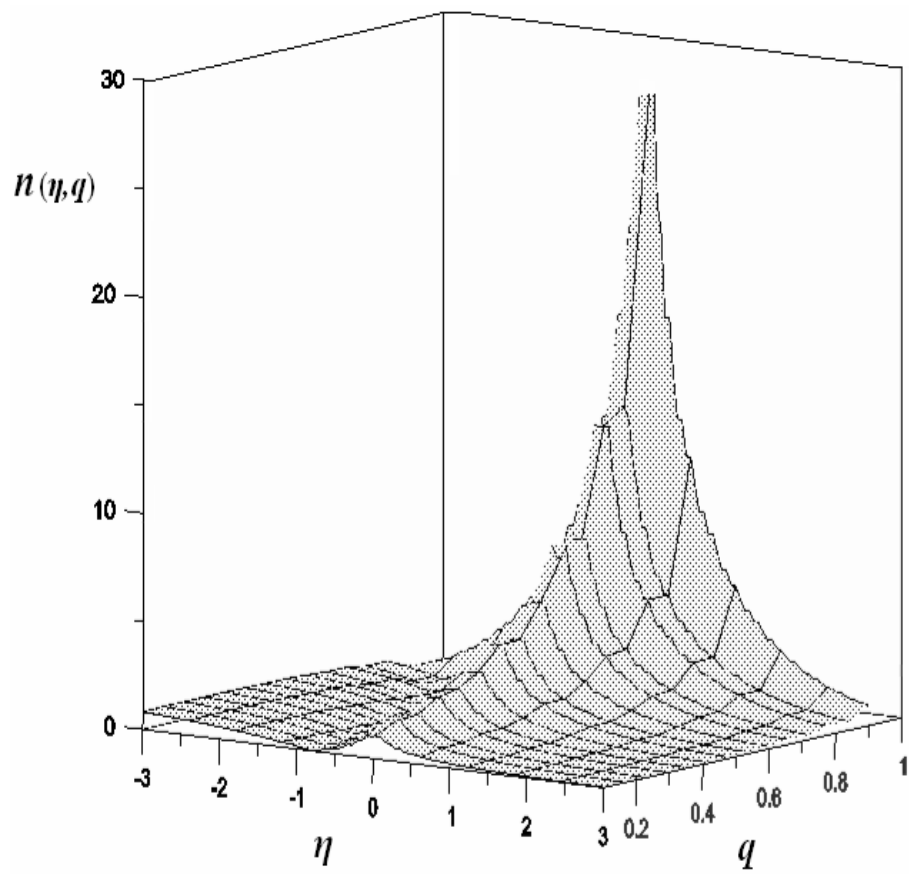

**FIG. 1.**

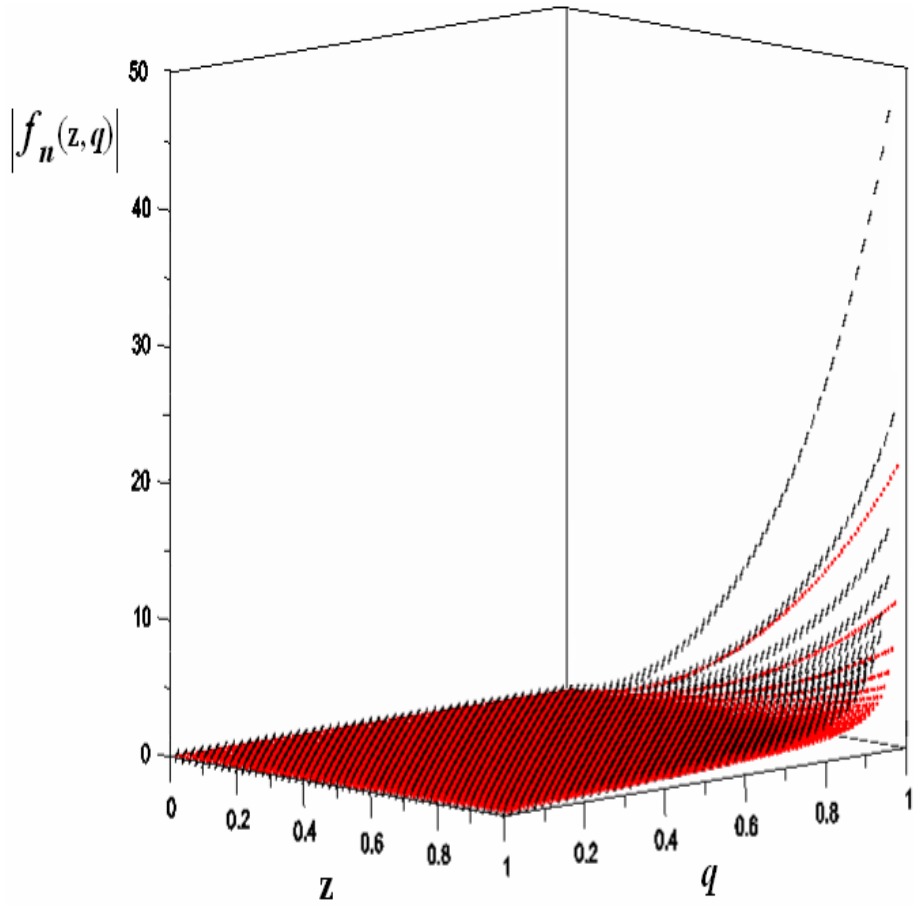

**FIG. 2.**

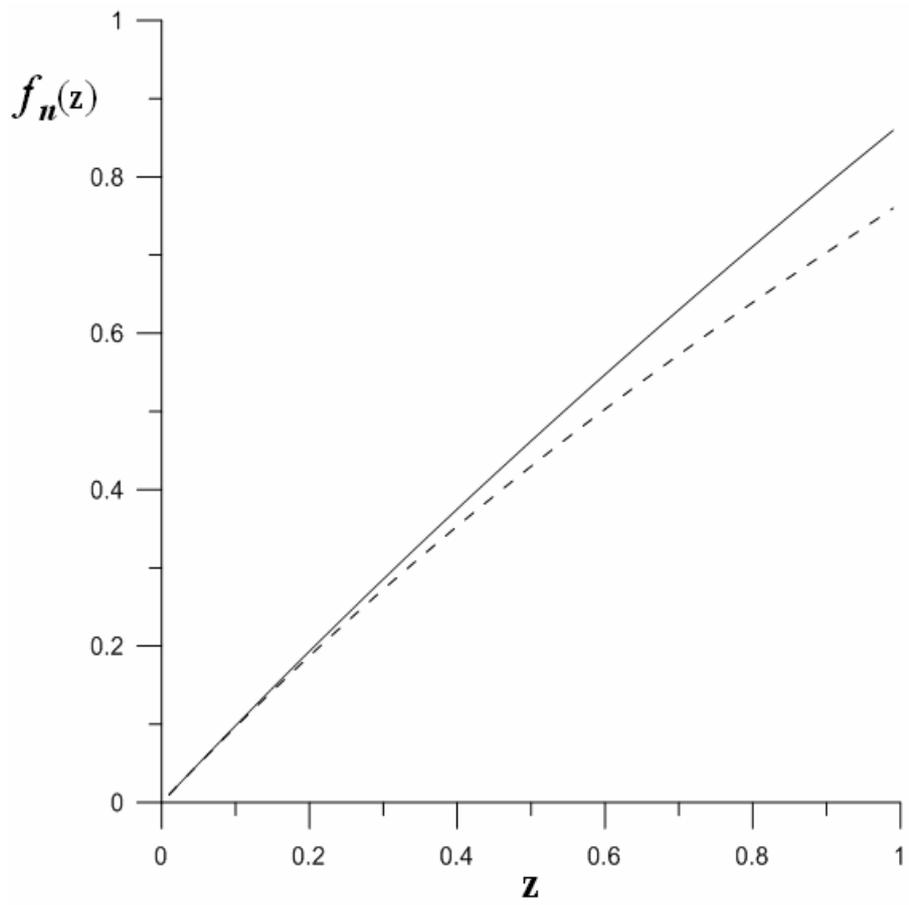

**FIG. 3.**

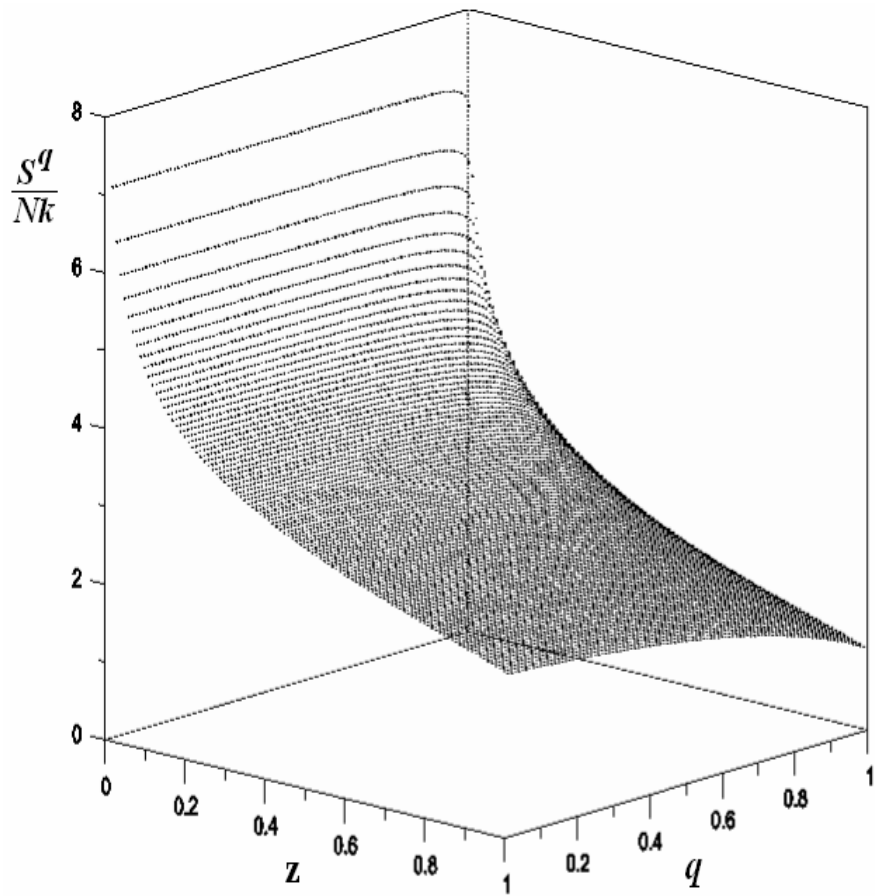

**FIG. 4.**

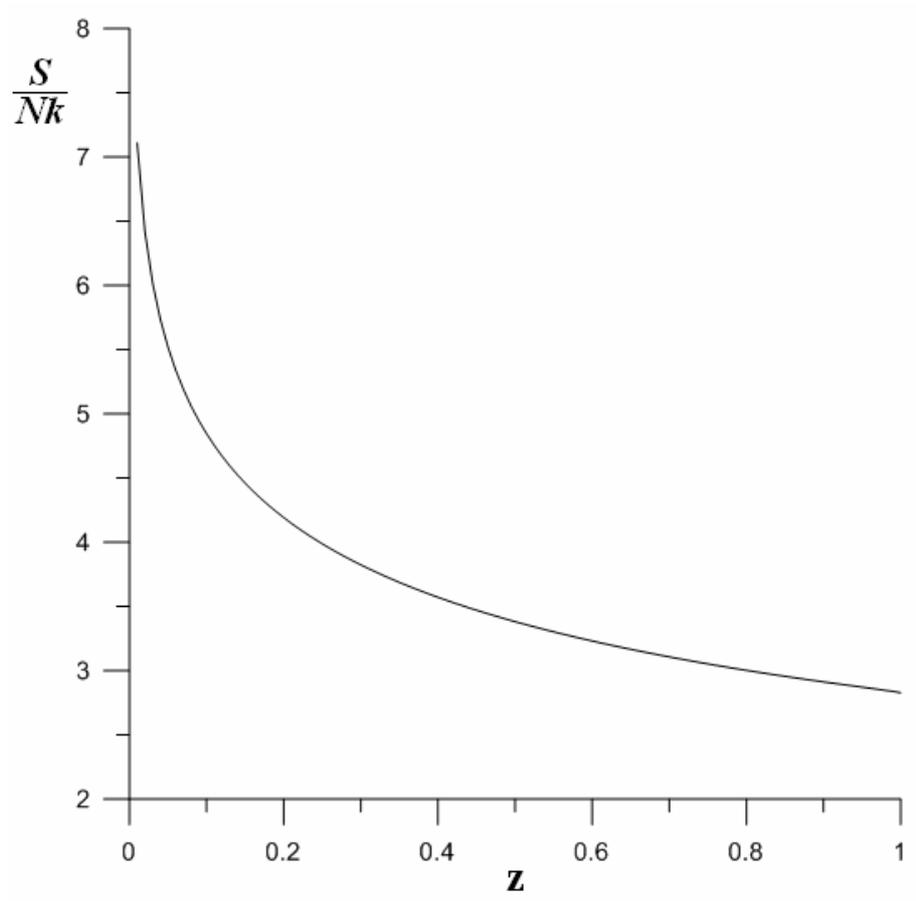

**FIG. 5.**

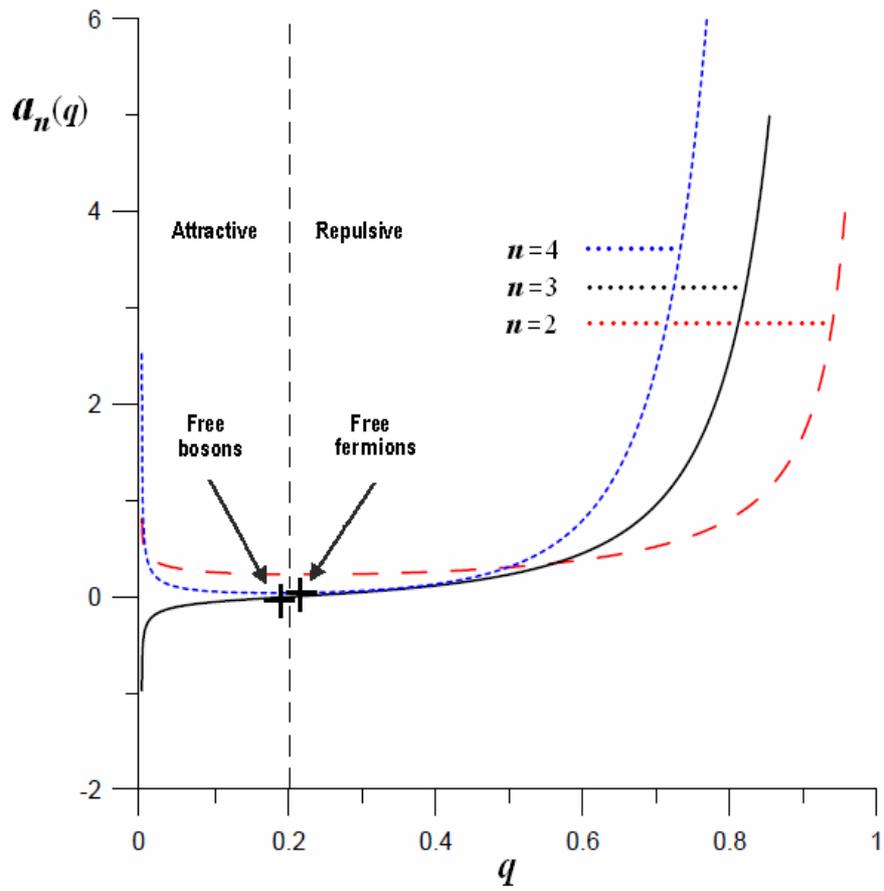

**FIG. 6.**

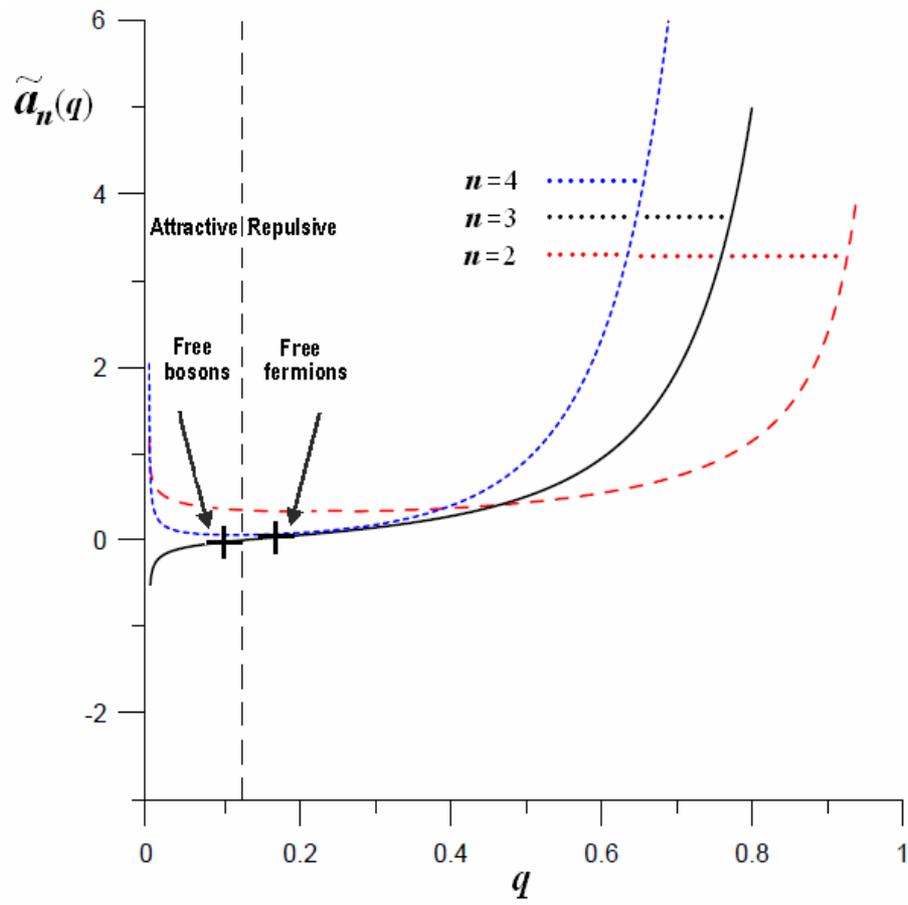

**FIG. 7.**